\documentclass{iau}
\usepackage{graphicx}

\title[Core Collapse Supernova Modeling] %% give here short title %%
{Toward Realistic Models of Core Collapse Supernovae: A Brief Review}

\author[Anthony Mezzacappa]   %% give here short author list %%
{Anthony Mezzacappa$^1$}

\affiliation{$^1$Department of Physics and Astronomy\\University of Tennessee, Knoxville\\Nielsen Physics Building -- 401\\1408 Circle Drive\\Knoxville, TN 37996-1200\\email: {mezz@utk.edu}}

\pubyear{2022}
\volume{362}  %% insert here IAU Symposium No.
\jname{Predictive Power of Computational Astrophysics\\ as a Discovery Tool}
\editors{D. Bisikalo, C. Boily, T. Hanawa, \& J. Stone, eds.}
\begin{document}

\maketitle

\begin{abstract}
Motivated by their role as the direct or indirect source of many of the elements in the Universe, numerical modeling of core collapse supernovae began more than five decades ago. Progress toward ascertaining the explosion mechanism(s) has been realized through increasingly sophisticated models, as physics and dimensionality have been added, as physics and numerical modeling have improved, and as the leading computational resources available to modelers have become far more capable. The past five to ten years have witnessed the emergence of a consensus across the core collapse supernova modeling community that had not existed in the four decades prior. For the majority of progenitors -- i.e., slowly rotating progenitors -- the efficacy of the delayed shock mechanism, where the stalled supernova shock wave is revived by neutrino heating by neutrinos emanating from the proto-neutron star, has been demonstrated by all core collapse supernova modeling groups, across progenitor mass and metallicity. With this momentum, and now with a far deeper understanding of the dynamics of these events, the path forward is clear. While much progress has been made, much work remains to be done, but at this time we have every reason to be optimistic we are on track to answer one of the most important outstanding questions in astrophysics: How do massive stars end their lives?
\keywords{supernovae: general, neutrinos, hydrodynamics, relativity}
\end{abstract}

\firstsection % if your document starts with a section,

\section{Introduction}
\label{sec:introduction}

Core collapse supernovae are directly or indirectly responsible for the lion's share of the elements in the Universe. As such, they are among the most important astrophysical phenomena to be studied and understood. There is a long history to core collapse supernova modeling, beginning with the work of \cite{CoWh66}. In the fifty six years since, such modeling has progressed considerably, shedding light on the key phenomena responsible for these explosions. This progress has been very encouraging to the core collapse supernova modeling community, but much work remains, as we will discuss here. 

\begin{figure}
\begin{center}
 \includegraphics[width=\textwidth]{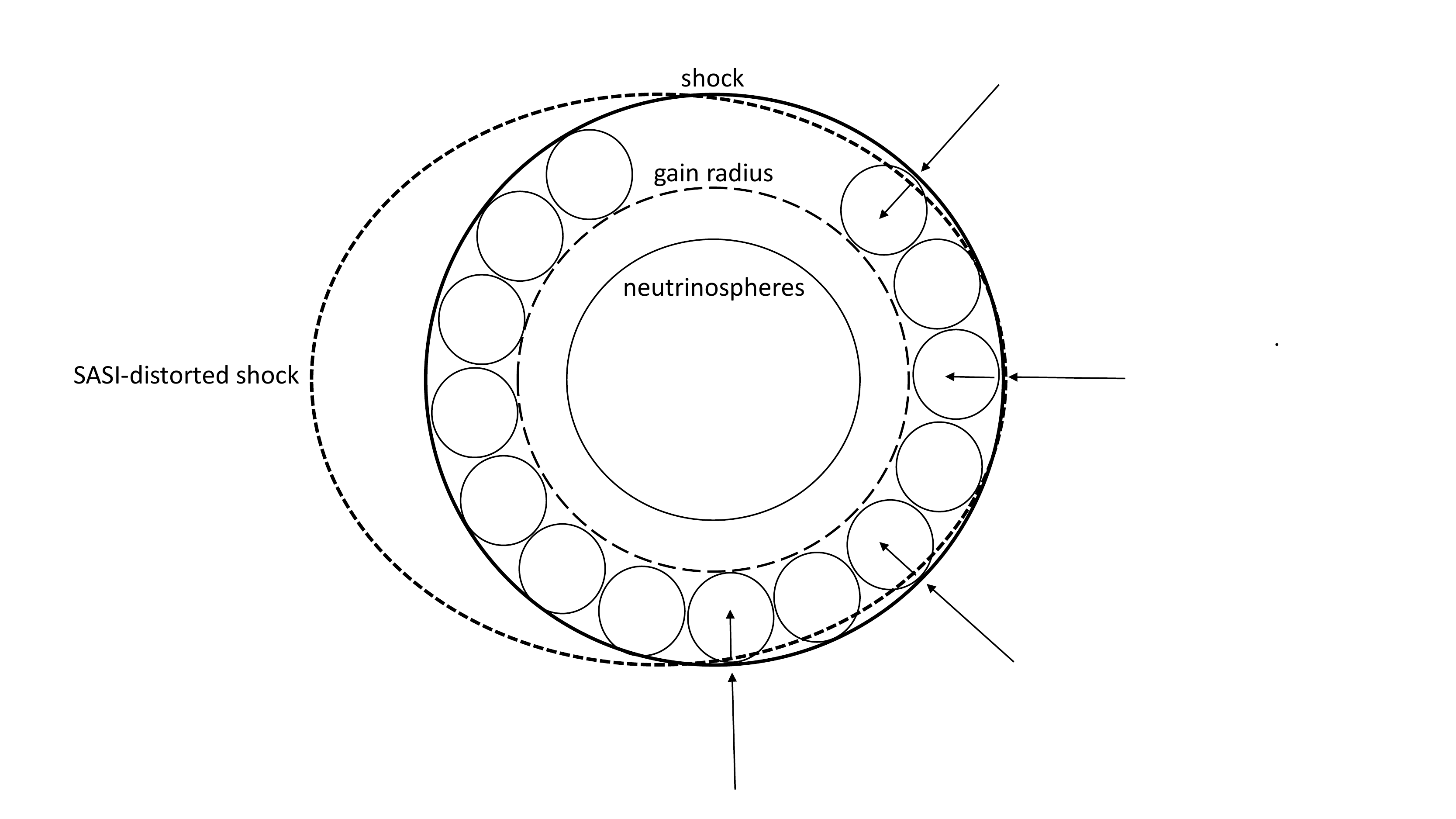} 
 \caption{After stellar core bounce and shock formation, the core is stratified into regions defined by the proto-neutrino star surface (neutrinospheres), the gain radius, and the stalled shock. The region between the neutrinospheres and the gain radius is a net neutrino cooling region. The region between the gain radius and the shock is a net neutrino heating region, also known as the gain region. Shown also, schematically, are convection in the gain region resulting from neutrino heating by the proto-neutron star below it, and the distortion of the shock due to the Standing Accretion Shock Instability (SASI).}
   \label{fig:postbounceconfig}
\end{center}
\end{figure}

Contemporary core collapse supernova theory (at least for the lion's share of such supernovae, which originate from slowly rotating progenitors) centers around the question of how the supernova shock wave, which forms as the result of stellar core collapse and bounce at super-nuclear densities and stalls as the result of the enervating processes of nuclear dissociation and neutrino losses, is revived. Oddly enough, the phenomena on which contemporary core collapse supernova theory rests entered the picture one at a time, about once every decade. While Colgate and White were the first to propose that core collapse supernovae could be neutrino driven, present-day efforts can be traced back to the work of Wilson in 1982, as documented in \cite{Wils85} and further elaborated in \cite{BeWi85}. Wilson demonstrated that
electron-neutrino and -antineutrino absorption on neutrons and protons below the shock, respectively, newly liberated from core nuclei by shock dissociation, can deposit sufficient energy to render the accretion shock a dynamical shock once again. Figure \ref{fig:postbounceconfig} illustrates the stratification of the stellar core region below the shock shortly after bounce. Neutrinos and antineutrinos of all three flavors emerge from the proto- neutron star. The region between the proto-neutron star surface and the shock divides into a net neutrino cooling region, due to the inverse of the weak interactions responsible for heating the material, and a net neutrino heating region above it. Neutrino heating and cooling balance at the gain radius. The gain region is defined as the region between the gain radius and the shock.
A decade later, core collapse supernova models were freed of the constraints imposed by spherical symmetry when the first two-dimensional simulations of \cite{HeBeCo92} were performed. Enter neutrino-driven convection. Not surprisingly, the post-shock material heated from below by neutrinos emerging from the neutrinospheres defining the surface of the proto-neutron star becomes convectively unstable, with multiple benefits with regard to aiding neutrino shock reheating. Most important, now continued accretion, which fuels the neutrino luminosities that drive the explosion, can continue while explosion develops. This is not possible in spherical symmetry. In 2003, yet another instability entered the picture. In the context of core-collapse-supernova-informed axisymmetric hydrodynamics studies, \cite{BlMeDe03} discovered that the supernova shock wave itself may become unstable to non-radial perturbations. This was confirmed in all subsequent axisymnmetric core collapse supernova simulations. In axisymmetry, the instability is manifest in an $\ell=1$ ``sloshing'' mode. In three dimensions, the sloshing mode is joined by an $m=1$ ``spiral'' mode (\cite{BlMe07}). As illustrated in Figure \ref{fig:postbounceconfig}, the SASI can increase the size of the gain region and, with it, post-shock neutrino heating. Neutrino-driven convection in the gain region becomes turbulent. In 2013, \cite{MuDoBu13} were the first to demonstrate quantitatively that the resulting turbulent ram pressure in this region assists the thermal pressure in the region in driving the shock outward. Rotation and magnetic field effects round out the list of phenomena that play a role in the explosion mechanism, though much of the progress we will discuss here was accomplished without them. Not surprisingly, the centrifugal effects of rotation, albeit slow rotation for the majority of massive stars, aid shock revival and explosion (e.g., see \cite{SuJaMe18}). And sufficient progress has been made to include magnetic fields in core collapse supernova models that we now know that magnetic effects are non-negligible, as well (e.g., see \cite{ObJaAl15}). Magnetic field strengths in the postshock region are amplified in the absence of rotation by collapse, convection, and turbulence, and by rotation when rotation is present, leading to magnetic stresses, like turbulent stresses, that aid in moving the shock outward to larger radii. The effects of magnetic fields become even more pronounced in the presence of significant core rotation, with outcomes changing qualitatively when magnetic fields are included (e.g., see \cite{KuArTa20}).

\section{Requirements}

More than two decades ago, \cite{BrDeMe01} and \cite{LiMeTh01} demonstrated, by comparing Newtonian and general relativistic simulation outcomes directly, that core collapse supernovae are general relativistic phenomena. Newtonian approximations to gravity, hydrodynamics (or magnetohydrodynamics if magnetic fields are included), and neutrino kinetics are, generally speaking, not realistic. With general relativistic treatments of all three, the stratification shown in Figure \ref{fig:postbounceconfig} is much more compact, with significantly reduced radii for the neutrinospheres, and gain and shock radii, giving rise also to a significantly reduced gain region volume. In addition, infall velocities ahead of the shock are larger. On the plus side of the ledger, the neutrinospheres are hotter, increasing the luminosities of the emergent neutrinos, as well as their RMS energies, though gravitational redshift will downgrade the neutrino spectra as they propagate outward through the cooling and heating layers.

The neutrino heating rate per gram in the gain region is given by

\begin{equation}
\dot{\epsilon}=\frac{X_n}{\lambda_{0}^{a}}\frac{L_{\nu_e}}{4\pi r^2} \left\langle E^{2}_{\nu_e} \right\rangle \left\langle \frac{1}{\mathcal{F}_{\nu_e}} \right\rangle
+\frac{X_p}{\bar{\lambda}_{0}^{a}}\frac{L_{\bar{\nu}_e}}{4\pi r^2} \left\langle E^{2}_{\bar{\nu}_e} \right\rangle \left\langle \frac{1}{\mathcal{F}_{\bar{\nu}_e}} \right\rangle,
\bigskip
\label{eq:heatingrate}
\end{equation}
where $\epsilon$ is the internal energy of the stellar core fluid per gram, $X_{n,p}$ are the neutron and proton mass fractions, respectively, $L_{\nu_e,\bar{\nu}_e}$ are the electron-neutrino and -antineutrino luminosities, respectively, $\mathcal{F}_{\nu_e,\bar{\nu}_e}$ are the inverse flux factors for the electron-neutrinos and -antineutrinos, respectively, and $\lambda_{0}^{a}, \bar{\lambda}_{0}^{a}$ are constants related to the weak interaction coupling constants. Thus, knowledge of the neutrino luminosities, spectra, and angular distributions are needed to compute the neutrino heating rates. This requires knowledge of the neutrino distribution functions, $f_{\nu_e,\bar{\nu}_e}(r,\theta,\phi,E,\theta_{p},\phi_{p},t)$, from which these quantities can be calculated. The neutrino distribution functions are determined by solving their respective Boltzmann kinetic equations. Thus, the core-collapse supernova problem is a phase space problem, in the end involving 6 dimensions plus time: 3 spatial dimensions (e.g., $r$, $\theta$, and $\phi$) and 3 momentum-space dimensions (as we will see: neutrino energy, a direction cosine, and an additional momentum-space angle).

It is not feasible at present, even on today's leadership-class supercomputing architectures, to perform multi-second (later we will see what time scales must be considered) core collapse supernova simulations with Boltzmann neutrino kinetics. Instead, today's leading core collapse supernova models are based on solutions to the equations for the neutrino angular moments, defined in terms of the neutrino distribution function. The moment equations are obtained by integrating the Boltzmann kinetic equations over neutrino angle. A two-moment approach is the canonical implementation, in which one solves for the spectral neutrino number or energy density and the spectral neutrino number or energy flux in each of three dimensions -- i.e., the lowest four angular moments of the neutrino distribution. Important angular information encoded by the neutrino distribution functions is kept, but not all of it. 
For example, the spectral number density and the three spectral number fluxes, one for each of the three spatial dimensions, are defined in terms of the distribution function as

\begin{equation}
\mathcal{N}(r,\theta,\phi,E,t)\equiv\int_{0}^{2\pi}d\phi_p\int_{-1}^{+1}d\mu f(r,\theta,\phi,\mu,\phi_p,E,t),
\label{eq:zerothmoment}
\end{equation}
\begin{equation}
\mathcal{F}^{i}(r,\theta,\phi,E,t)\equiv\int_{0}^{2\pi}d\phi_p\int_{-1}^{+1}d\mu n^{i}f(r,\theta,\phi,\mu,\phi_p,E,t),
\bigskip
\label{eq:firstmoment}
\end{equation}
respectively, where $\mu\equiv\cos\theta_p$ is the neutrino direction cosine defined by $\theta_p$, one of the angles of propagation defined in terms of the outward pointing radial vector defining the neutrino's position at time $t$. In three dimensions, two angles are needed to uniquely define a neutrino propagation direction. The angle $\phi_p$ provides the second. $n^i$ is the component of the neutrino direction cosine in the $i^{\rm th}$ direction, given as a function of $\mu$ and $\phi_p$. $E$ is the neutrino energy. $E,\theta_p,\phi_p$ can be viewed as spherical momentum-space coordinates. Integration of the neutrino Boltzmann equation over the angles $\theta_p$ and $\phi_p$, weighted by $1$, $n^i$, $n^{i}n^j$, ... defines an infinite set of evolution equations for the infinite number of angular moments of the distribution function, which is obviously impossible to solve. In a moments approach, the infinite set of equations is rendered finite by truncation, after the equation for the zeroth moment in the case of one-moment closure (e.g., flux-limited diffusion) or after the equations for the first moments in the case of two-moment closure (e.g., M1 closure). Neutrinos are, of course, Fermions, and Fermions obey Fermi--Dirac statistics. As a result, closure of the system of moment equations must also obey Fermi--Dirac statistics. It must be {\em realizable}. But not all of the closures implemented in leading core-collapse supernova models to date are. Commonly used closures such as the Minerbo, Levermore, and Cernohorsky--Bludman closures obey Maxwell--Boltzmann, Bose--Einstein, and Fermi--Dirac statistics, respectively, with the Minerbo closure being the one most commonly deployed. This does not present an issue at low occupancies. Rather, it is in the important region deep within the proto-neutron star, where neutrino occupancies are large and their fluxes are small, where we can expect a breakdown of realizability (see \cite{ChEnHa19}).

The coupling of the neutrinos to the stellar core matter is mediated by what is now known to be an extensive set of weak interactions. Charged-current electron-neutrino and -antineutrino absorption on nucleons behind the shock powers the supernova, but the neutrino luminosities and RMS energies emerging from the neutrinospheres, which enter the heating rate, Equation (\ref{eq:heatingrate}), are defined by a much larger set of charged- and neutral-current interactions -- critical among these: non-isoenergetic scattering on nucleons, electron--positron annihilation, and nucleon--nucleon bremsstrahlung. For a complete listing, the reader is referred to \cite{MeEnMe20}. As we will discuss later, the set of relevant neutrino weak interactions continues to evolve. With regard to the the neutrino weak interactions, two things must be emphasized: (1) All of the relevant neutrino interactions in the above-cited list must be included. (2) State of the art implementations of each of the interactions, as described in \cite{MeEnMe20}, must be used.

\section{The Current State of the Art}
\label{sec:stateoftheart}
\begin{figure*}
%\begin{center}
\includegraphics[width=0.5\textwidth]{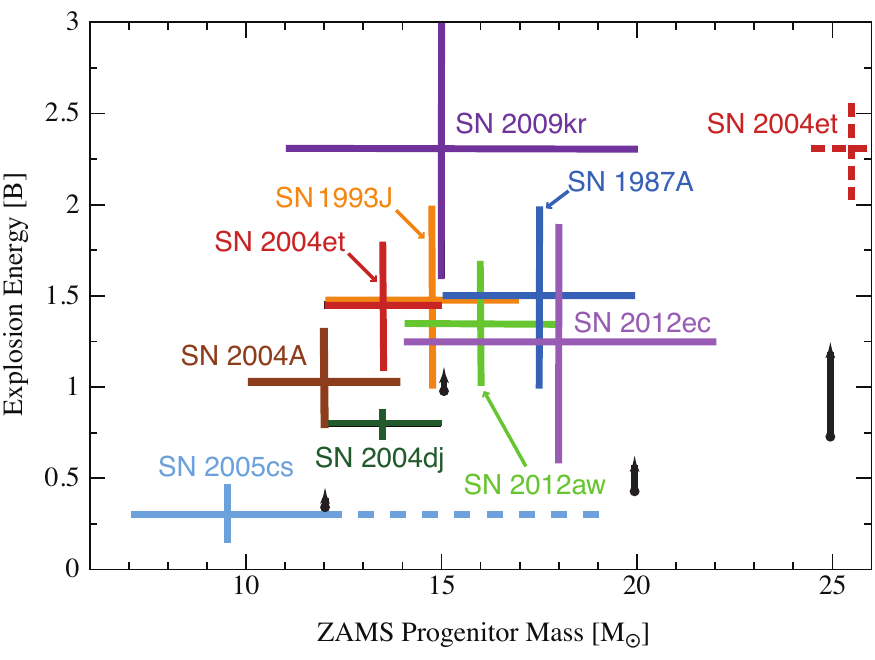}
\includegraphics[width=0.5\textwidth]{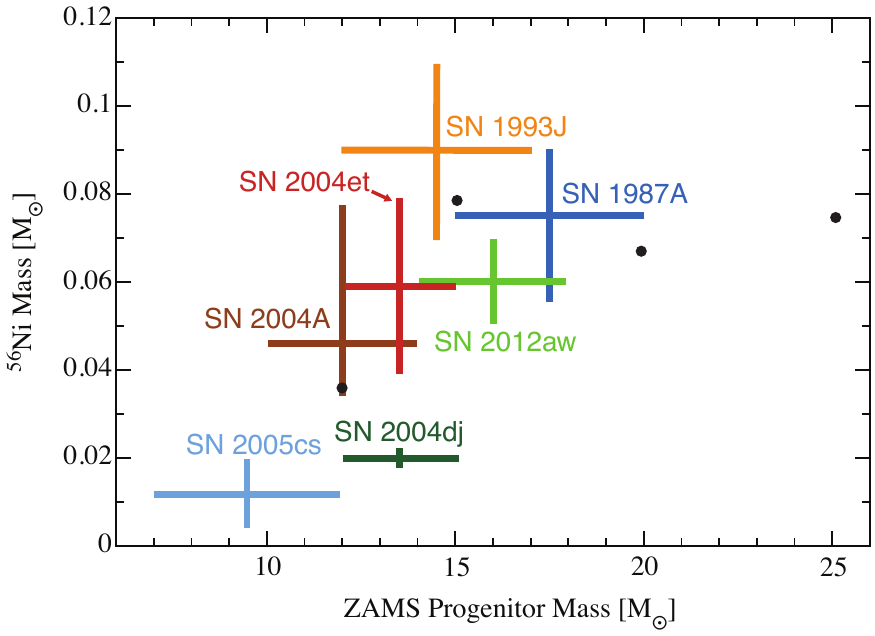}
\caption{Explosion energies and $^{56}$Ni mass produced (black dots) for four axisymmetric core collapse supernova models for progenitors of 12, 15, 20, and 25 M$_\odot$, plotted against observations as a function of ZAM mass (from \cite{BrLeHi16}). The black arrows in the explosion energies plot indicate that the explosion energies are still increasing at the times the simulations were stopped. The length of the arrows is tied to the magnitude of the rates of change of the explosion energies at these times.}
\label{fig:explosionenergies}
%\end{center}
\end{figure*}

Despite the challenges the core collapse supernova modeling community has faced and that, to a certain extent, still lie ahead, dictated by the rather austere requirements just discussed, we are arguably in a very encouraging period of core collapse supernova theory. At present, the efficacy of neutrino shock revival has been demonstrated by all leading core collapse supernova modeling groups worldwide, across progenitor characteristics such as mass, metallicity, and rotation (\cite{HaMuWo13,LeBrHi15,MeJaMa15,MeJaBo15,SuJaMe18,RoOtHa16,KuTaKo16,OcCo18,VaBuRa19,BuRaVa19,KuArTa20,StJaKr20}). This is a marked change relative to where the field was only ten years ago, let alone more than fifty years ago. The modeling community is now entering a new period, of quantitative rather than qualitative prediction. For example, modelers are now able to compare quantitatively predictions for observables such as explosion energies, $^{56}$Ni production, and remnant neutron star masses and kicks, with observations. An example of this is shown in Figure \ref{fig:explosionenergies}, where the explosion energy predictions of \cite{BrMeHi13} and \cite{BrLeHi16} for axisymmetric simulations beginning with progenitors of mass 12, 15, 20, and 25 M$_\odot$ are plotted against observed explosion energies, as a function of progenitor mass. A more recent example of this encouraging agreement, across a large collection of progenitor masses, can be found in \cite{BuVa21}.

The above cited efforts provide a strong foundation on which to build future core collapse supernova models. Nonetheless, they suffer from one or more of three significant shortcomings: (1) General relativity is included only approximately. (2) Neutrino transport is not three-dimensional. (3) They do not include all of the necessary neutrino interactions and/or state of the art treatments of them. We consider each of these in turn. 

The models of \cite{HaMuWo13,LeBrHi15,MeJaMa15,MeJaBo15,SuJaMe18,OcCo18,VaBuRa19,BuRaVa19,StJaKr20} all deploy an effective gravitational potential in the context of Newtonian hydrodynamics. In this approach, the monopole contribution to the Newtonian gravitational potential, as defined through a decomposition of the potential in terms of spherical harmonics, is replaced by a corrected potential whose form is suggested by comparing the equations for Newtonian hydrostatic equilibrium and the Tolman--Oppenheimer--Volkov (TOV) equation for general relativistic hydrostatic equilibrium. At an instant of time in the simulation, the three-dimensional data is spherically averaged, and the resulting quantities are used to compute the corrected potential. This approximation to general relativistic gravity was first proposed by \cite{RaJa02}. The goal is to capture the stronger gravitational fields that result in general relativity given that, along with rest mass, internal energy and pressure contribute to the gravitational field, as well. The fundamental problem with this approach is, of course, that the correction for general relativity is {\em ad hoc}. In the context of two-dimensional models, \cite{MuJaMa12} obtained qualitatively different results (e.g., explosion versus no explosion) in two models of the same progenitor mass, with Newtonian hydrodynamics and an effective potential used in one model and general relativistic hydrodynamics and gravity used in the other. Explosion occurred in the general relativistic model.

\begin{figure}
\begin{center}
\includegraphics[width=0.6\textwidth]{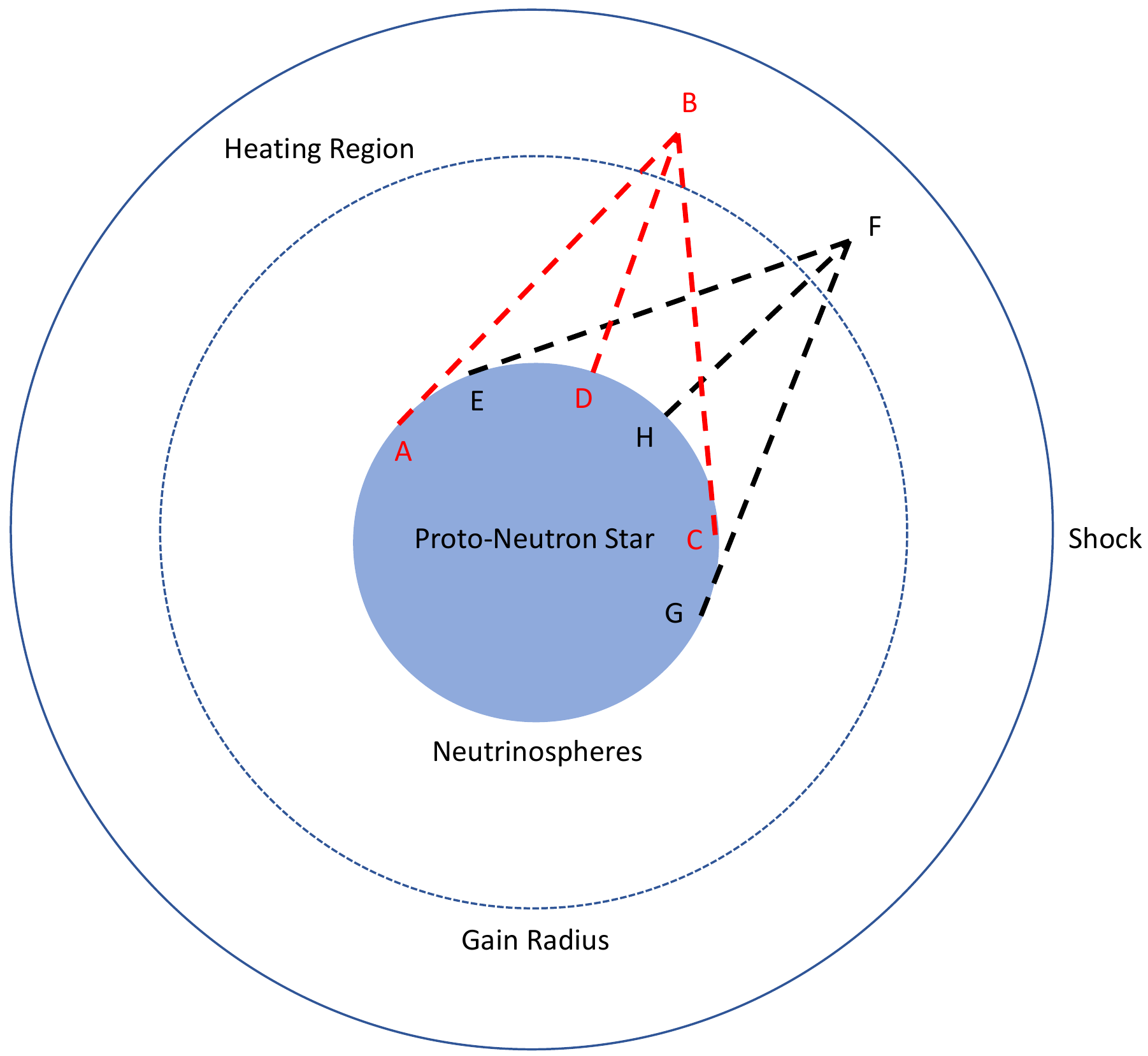}
\caption{In the ray-by-ray approximation, a hot spot at D would correspond to a hot surface between A and C, thereby overestimating the neutrino heating at B. A (relatively) cold spot at H would correspond to a (relatively) cold surface between E and G, thereby underestimating the heating at F, to which the hot spot at D would in reality contribute.}
\label{fig:RbR}
\end{center}
\end{figure}

The models of \cite{HaMuWo13,LeBrHi15,MeJaMa15,MeJaBo15,SuJaMe18,StJaKr20} deploy the so-called Ray-by-Ray (RbR) approximation to neutrino transport. The sophistication of neutrino transport implementations in spherically symmetric core collapse supernova models motivated this approach, which was first proposed by \cite{RaJa02}. For each $(\theta,\phi)$ on an angular grid in a numerical simulation, a solution of the neutrino transport equations is obtained for all $r$, assuming spherical symmetry. For a spherically symmetric source (in our case, given a spherically symmetric proto-neutron star), the RbR solution is exact. If the conditions at the neutrinospheres differ from spherical symmetry over a protracted period of time, the RbR approximation will, as a function of $(\theta,\phi)$, lead to over- and under-estimation of the neutrino heating in the gain region (see Figure \ref{fig:RbR}). Such protracted deviations from spherical symmetry can happen in, for example, axisymmetric simulations. Given the imposed symmetry, accretion funnels impinging on the neutrinospheres, thereby heating them, can be pinned to particular $(\theta,\phi)$, leading to inaccurate neutrino heating, which in turn can alter the explosion outcome artifically (e.g., see \cite{SkBuDo16}). More recently, the RbR approximation was investigated by \cite{GlJuJa19} in the context of three-dimensional models and, given the absence of any imposed symmetry, yielded results in good agreement with three-dimensional transport, in the cases considered. Nonetheless, future simulations should endeavor to deploy three-dimensional neutrino transport.

The models of \cite{RoOtHa16,KuTaKo16,KuArTa20}, while three-dimensional and general relativistic (although \cite{RoOtHa16} neglect velocity-dependent terms in the neutrino moment equations), deploy a subset of the relevant neutrino interactions and, in some cases, not the most advanced implementations of them. Specifically, in all three cases, the neutrino opacities given by \cite{Brue85} are used. In the simulations conducted by \cite{KuTaKo16,KuArTa20}, the production of neutrino--antineutrino pairs via nucleon--nucleon bremsstrahlung, as given by \cite{HaRa98}, is used, as well. In the \cite{Brue85} opacities, electron capture on nuclei is treated using the Independent Particle Model (IPM) for nuclei. In this model the nucleons in the nuclei are assumed to be noninteracting. Given the IPM, electron capture on nuclei during stellar core collapse is blocked, due to neutron final-state blocking in this approximation, and electron capture is given by capture on free protons only. Consequently, the deleptonization of the core (the electron neutrinos produced by electron capture initially escape until neutrino trapping densities are reached) is underestimated, which in turn changes the initial shock location and strength. Simulations by \cite{HiMeMe03}, which used the more advanced models of electron capture on nuclei of \cite{LaMaSa03}, in which interactions between nucleons in nuclei are taken into account, as well as thermal effects, demonstrated that the shock forms more deeply in the core, is less energetic when it does form, and has to propagate through more of the iron core before reaching the silicon layer and the density drop-off that accelerates explosion. \cite{Brue85} opacities also assume that scattering on nucleons is isoenergetic. While the large rest mass of the nucleons leads kinematically to small neutrino energy transfer, \cite{MuJaMa12} demonstrated that the high collision rate in the vicinity of the neutrinospheres leads to heating of the electron-neutrinospheres and, in turn, neutrino shock reheating itself. They deployed the more advanced neutrino--nucleon scattering rates of \cite{BuSa98} and \cite{RePrLa98}, which account for the small but important energy exchange that occurs in these scattering events. Finally, \cite{Brue85} opacities include the production of neutrino--antineutrino pairs via electron--positron annihilation but neglect the additional source of neutrino pairs from nucleon--nucleon bremsstrahling, which can dominate neutrino pair production via electron--positron annihilations in certain regions of the core.

\section{Additional Challenges}
\subsection{Time Scales}
The significant rates of change in the explosion energies in some of the explosion models considered in Figure \ref{fig:explosionenergies} are an indication that explosion energies in multi-dimensional models will evolve over much longer time scales than the canonical one-second of postbounce evolution considered in the past. In fact, the compendium of two-dimensional models published to date indicate that several seconds of postbounce simulation will be required (e.g., see \cite{BuVa21}). While not so much a challenge for two-dimensional simulations, three-dimensional simulations are much more costly. Among the dozens of explosion simulations documented in the references cited in this review, the final explosion energies have been determined in only a few cases, for this reason. This illustrates the general problem we face in conducting a sufficient number of three-dimensional core collapse supernova models to span progenitor parameter space while at the same time conducting such simulations for sufficiently long periods of time in order to predict all relevant observables and compare these predictions with observations. At this time, advances in core collapse supernova theory are throttled more by the availability of computing time than by the necessary advances documented in Section \ref{sec:stateoftheart}.

\subsection{Progenitors}
Much of the progress in core collapse supernova theory has been achieved by considering spherical, non-rotating progenitors of various masses and metallicities. The reason for this is simple. There are at present no core collapse supernova progenitors that have been obtained through three-dimensional stellar evolution simulations, and there will not be any such progenitors for some time to come. Nonetheless, important progress has been made to determine the impact such simulations and the progenitors they will produce may have on core collapse supernova models. There are two primary considerations, based on considerations of single-star and binary-star evolution: (1) What deviations from spherical symmetry at the onset of collapse might we expect and what impact will those deviations have on the supernova mechanism? (2) What is the impact on the evolution of the progenitor of being in a binary system?

\cite{CoChAr15} were the first to consider the impact of deviations from spherical symmetry for single massive stars. They followed the final minutes of silicon burning in three dimensions, to the onset of core collapse and, subsequently, followed collapse, bounce, shock formation, and shock propagation. The non-spherical progenitor structure in their model induced by convective silicon burning led to enhanced post-shock turbulence, which in turn aided explosion. That deviations from spherical symmetry in core collapse supernova progenitors aids explosion was corroborated in the later work of \cite{MuMeHe17} and \cite{VaCoBu21}. Going forward, models in three dimensions should begin, at least in some cases for a comparative analysis, with the late stages of stellar evolution rather than with the onset of collapse, until three-dimensional progenitors evolved through all stages of stellar evolution become available. Of course, stellar evolution simulations will face the same challenges. It will be difficult to perform the number of three-dimensional stellar evolution simulations needed to adequately span the full range of core collapse supernova progenitors.

SN1987A occupies a special place in core collapse supernova theory. The very recent work of \cite{UtWoJa21} determined that only one of the three-dimensional models considered, a binary progenitor model, satisfied (almost) all of the observational constraints. Also in the context of three-dimensional models, earlier efforts to understand the impact of binary stellar evolution on the progenitors of core collapse supernovae -- specifically, the impact of mass loss in binary systems -- were initiated by \cite{MuTaHe19}. They considered both single-star progenitors between 9.6 and 12.5 M$_\odot$ and ultra-stripped progenitors with He core masses between 2.8 and 3.5 M$_\odot$. They concluded that the differences between core collapse supernova models initiated from single or binary-stripped stars {\em of the same He core mass} were no larger than the stochastic variations among models initiated from single stars. However, they did stress the importance of binary evolution on the determination of the He core mass itself. A second, complementary study by \cite{VaLaRe21} investigated the impact of binary mass loss during the first Roche-lobe overflow phase, considering the impact on stripped but not ultra-stripped progenitors. They concluded that, for the same initial mass, binary-stripped progenitors are more ``explodable'' relative to their single-star counterparts. The results of these studies make clear that at least some future three-dimensional core collapse supernova models must factor in the effects of binary stellar evolution on the progenitors used.

\subsection{Weak Interactions}
The early history of core collapse supernova theory was intertwined with the development of the electroweak theory of the weak interactions (see \cite{MeEnMe20}).  As weak interaction theory progressed, so did core collapse supernova theory. This continues to this day. There are two things to consider: (1) The continued addition of new weak interaction channels of relevance to the core collapse supernova mechanism. (2) The uncertainties associated with the cross sections for all of the weak interactions involved. We discuss both here.

The work by \cite{BoJaLo17} is the most recent example of the addition of a whole new class of weak interaction channels to core collapse supernova theory. They demonstrated that muons cannot be ignored as a constituent of the stellar core material and, with them, neutrino--muon interactions. Past simulations ignored muons given their large rest mass and, consequently, their small expected populations, but the work cited here makes clear this is not a good assumption. \cite{BoJaLo17} found that the inclusion of muons could {\em qualitatively} alter the outcome of the supernova simulations they performed.

On the other hand, the work by \cite{MeJaBo15} considered the sensitivity of core collapse supernova models to {\em uncertainties} in the cross sections of the weak interactions already included in all leading core collapse supernova models. In particular, they considered variations of the neutrino--nucleon scattering cross section consistent with experimental uncertainties and found that variations of this cross section could again qualitatively change the outcome of the models they considered. Neutrino--nucleon scattering is one of the most important opacities in the proto-neutron star. This motivated the decision to vary this particular opacity. A more systematic and complete study will need to be undertaken in which all of the neutrino opacities are varied in a statistically meaningful way. Such a study would take into consideration the known feedbacks that occur when neutrino opacities are varied (e.g., see \cite{LeMeBr12}). Unfortunately, in the context of three-dimensional modeling, such sensitivity studies are at present prohibitive. Going forward, at least in the context of three-dimensional modeling, we will need to rely on more limited studies based on varying targeted opacities, as was done in the studies cited here.

\section{Horizons}
The core collapse supernova simulations cited here, which represent the culmination of decades of core collapse supernova modeling, assumed that neutrinos are massless. We now know this is not the case. Neutrinos have mass and, as a result, quantum mechanical effects such as vacuum flavor mixing, matter-enhanced flavor mixing, and finally, and perhaps most importantly, neutrino-enhanced flavor mixing can occur. Such effects can be expected to impact observations of core collapse supernova neutrinos from the next Galactic or near-extra-Galactc event, are likely to impact core collapse supernova nucleosynthesis, and may even impact the explosion mechanism itself. Here, we concern ourselves with the last possibility.
 
Electron neutrinos and antineutrinos interact via both charged and neutral currents, whereas muon and tau neutrinos and antineutrinos interact via neutral currents only. As a result, muon and tau neutrinos decouple from the stellar core material at higher densities and, consequently, higher temperatures, and emerge with harder spectra. On the other hand, neutrino shock revival is mediated by electron neutrinos and antineutrinos. Mixing between electron flavor and muon and tau flavor neutrinos, if it occurs below the gain region, could enhance neutrino shock reheating given its sensitive dependence on the electron flavor neutrino and antineutrino spectra [see Equation (\ref{eq:heatingrate})]. Herein lies the importance of neutrino mixing to the explosion mechanism.

Core collapse supernovae are unique neutrino environments. Nowhere else in the Universe do we have trapped degenerate seas of neutrinos and antineutrinos of all flavors, diffusing through the proto-neutron star, emerging from the region in which the neutrinospheres are embedded. In such environments, neutrino--neutrino interactions become important. Of particular note here: 
\cite{Sawyer05} demonstrated that under certain conditions, which depend on the angular distributions of the electron neutrinos and antineutrinos in the core, so-called ``fast'' flavor transformations can occur in the neutrino decoupling region around the neutrinospheres. Given these would occur in regions below the gain region, they would impact neutrino shock reheating -- i.e., they would factor into the core collapse supernova explosion mechanism. Indeed, in the context of three-dimensional models, \cite{AbCaGl21} and \cite{NaBuJo21} found  conditions for fast flavor transformations in the post shock region in all of the models they considered that exploded. Given that fast flavor transformations and explosion coexist in these models, the models' outcomes (e.g., the final explosion energies) may depend on fast-flavor-transformation physics. For a review of this rapidly evolving subject, the reader is referred to \cite{DuFuQi10}, \cite{Miri16}, and \cite{TaSh21}.

To include flavor transformations in core collapse supernova simulations, classical neutrino kinetics -- specifically Boltzmann neutrino kinetics, given the need to keep track of the neutrino and antineutrino angular distributions -- would need to be replaced by quantum neutrino kinetics. A three-dimensional, general relativistic treatment of quantum neutrino kinetics, with sufficient angular and energy resolution in neutrino momentum space and a complete set of neutrino weak interactions, remains a long-term goal.

\section{Summary and Outlook}
Efforts to ascertain the core collapse supernova explosion mechanism(s) have been ongoing for more than five decades. Recent, rapid progress is very promising. Based on the sophisticated three-dimensional simulations performed to date, which span progenitor mass and metallicity, the majority with no or slow rotation, there is consensus across core collapse supernova modeling groups that core collapse supernovae can be driven by neutrino shock reheating aided by turbulent neutrino-driven convection and, depending on the progenitor, the SASI. Nonetheless, much work remains to be done. First and foremost, all three-dimensional models with classical neutrino kinetics must be further developed to include (1) general relativity for gravity, hydrodynamics, and neutrino kinetics, (2) magnetohydrodynamics to capture the role magnetic fields play for both slowly and rapidly rotating progenitors, and (3) the full set of relevant neutrino weak interactions. In the long-term, efforts to extend classical neutrino kinetics to quantum neutrino kinetics must continue, to explore the role of fast flavor transformations. This quantum kinetics development must ultimately be founded on the extension of classical neutrino kinetics based on the moment formalism, used in most of the three-dimensional models cited here, to classical neutrino kinetics based on solutions of the neutrino Boltzmann equations. At the same time, all models must remain abreast of the evolving set of relevant neutrino weak interactions and the increasingly constrained set of viable nuclear equations of state (e.g., see \cite{TeLaOh17}).

In the past, the core collapse supernova modeling community has contended with uncertainties associated with (a) approximations to critical components of core collapse supernova models,  (b) differences in the numerical methods deployed by different groups, (c) the differing numerical resolutions used in the simulations carried out by these groups, (d) the neutrino interaction cross sections, and (e) the nuclear equation of state. Moreover, the past five decades has taught us that core collapse supernova models are sensitive to small changes in important quantities, such as the neutrino luminosities, RMS energies, cross sections, etc. While this may seem daunting at first, the core collapse supernova modeling community has made significant progress over the past fifty-plus years through systematic improvements that address all of the categories of uncertainty listed above, and no doubt will continue to do so in the future.

For additional information, we refer the reader to recent reviews by \cite{JaMesu16}; \cite{Mueller16}; \cite{Mueller20}; \cite{MeEnMe20}; and \cite{BuVa21}.
\bigskip

AM acknowledges support from the National Science Foundation through grants PHY 1806692 and 2110177, the Department of Energy through its Scientific Discovery through Advanced Computing Program through grant DE-SC0018232, and the Department of Energy's Exascale Computing Project.

\end{document}